\documentclass{emulateapj}
\submitted{ApJ, in press}
\newcommand{\kms}{\,km\,s$^{-1}$}
\newcommand{\ergs}{\,erg\,s$^{-1}$}
\newcommand{\ergcm}{\,erg\,cm$^{-2}$\,s$^{-1}$}

\newcommand{\cxo}{\emph{Chandra}}
\newcommand{\xmm}{\emph{XMM-Newton}} 
\newcommand{\nh}{$N_{\rm H}$}
\newcommand{\cms}{\,cm$^{-2}$}
\newcommand{\psr}{PSR J1119$-$6127}
\newcommand{\snr}{G292.2$-$0.5}
\newcommand{\pasa}{PASA}

\begin{document}

\title{Deep X-ray Observations of the Young High-Magnetic-Field Radio Pulsar
J1119$-$6127 and Supernova Remnant \snr}
\author{C.-Y. Ng\altaffilmark{1}, V. M. Kaspi\altaffilmark{1}, W. C. G.
Ho\altaffilmark{2}, P. Weltevrede\altaffilmark{3}, S.
Bogdanov\altaffilmark{4}, R. Shannon\altaffilmark{5}, and M. E.
Gonzalez\altaffilmark{6}}
\email{ncy@physics.mcgill.ca}

\altaffiltext{1}{Department of Physics, McGill University, Montreal, QC H3A 2T8, Canada} 
\altaffiltext{2}{School of Mathematics, University of Southampton, Southampton, SO17 1BJ, UK}
\altaffiltext{3}{Jodrell Bank Centre for Astrophysics, The University of Manchester, Alan Turing Building, Manchester M13 9PL, UK}
\altaffiltext{4}{Columbia Astrophysics Laboratory, Columbia University, 550 West 120th Street New York, NY 10027, USA}
\altaffiltext{5}{CSIRO Astronomy and Space Sciences, Australia Telescope National Facility, Marsfield, NSW 2210, Australia}
\altaffiltext{6}{Department of Physics and Astronomy, University of British Columbia, Vancouver, BC V6T 1Z1, Canada}

\shorttitle{Deep X-ray Observations of \psr\ and SNR \snr}
\shortauthors{Ng et al.}

\begin{abstract}
High-magnetic-field radio pulsars are important transition objects for
understanding the connection between magnetars and conventional radio pulsars.
We present a detailed study of the young radio pulsar J1119$-$6127, which has
a characteristic age of 1900\,yr and a spin-down-inferred magnetic field of
$4.1\times10^{13}$\,G, and its associated supernova remnant \snr, using deep
\emph{XMM-Newton} and \emph{Chandra X-ray Observatory} exposures of over
120\,ks from each telescope. The pulsar emission shows strong modulation below
2.5\,keV, with a single-peaked profile and a large pulsed fraction of
$0.48\pm0.12$. Employing a magnetic, partially ionized hydrogen atmosphere
model, we find that the observed pulse profile can be produced by a single hot
spot of temperature 0.13\,keV covering about one third of the stellar surface,
and we place an upper limit of 0.08\,keV for an antipodal hot spot with the
same area. The nonuniform surface temperature distribution could be the result
of anisotropic heat conduction under a strong magnetic field, and a
single-peaked profile seems common among high-$B$ radio pulsars. For the
associated remnant \snr, its large diameter could be attributed to fast
expansion in a low-density wind cavity, likely formed by a Wolf-Rayet
progenitor, similar to two other high-$B$ radio pulsars.
\end{abstract}


\section{Introduction}
Over the past two decades, our understanding of neutron stars has been
revolutionized due to discoveries of several new classes of objects
\citep[see][for a review]{kas10}. An extreme class is magnetars, which
typically have high spin-down-inferred magnetic fields\footnote{The dipole
$B$-field can be estimated by $B=3.2\times10^{19}(P\dot P)^{1/2}$\,G, where
$P$ is the spin period in second and $\dot P$ is the spin-down rate.} of
$10^{14}$--$10^{15}$\,G and show violent bursting activities
\citep[see][]{re11,mer08,nkd+11,snl+12}. It is generally believed that their
X-ray emission is powered by the decay of ultra-strong magnetic fields
\citep{dt92,td95,td96}. However, the origin of magnetars and their relation to
the more conventional rotation-powered pulsars remains a puzzle. The
distinction between these two classes of objects became increasingly blurred
thanks to the recent discoveries of a relatively low $B$-field magnetar of $B<
7.5\times10^{12}$\,G \citep{ret+10}, magnetar-like bursts from a
rotation-powered pulsar \citep{ggg+08,nsg+08,ks08}, and radio emission from
magnetars \citep{crh+06,crh+07,lbb+10}. These provide some hints that
magnetars and radio pulsars could be drawn from the same population, and the
former could represent the high-field tail of a single birth $B$-field
distribution \citep{km05,pp11a,ho12}. One way to test this ``unification'' picture
is via observations of high-magnetic-field radio pulsars,\footnote{Throughout
this paper, we refer to rotation-powered pulsars as radio pulsars, even though
their radio beams may miss the Earth, e.g.\ PSR J1846$-$0258 \citep{akl+08}.}
which are a critical group of rotation-powered pulsars that have similar spin
properties as magnetars, implying magnetar-like dipole $B$-field strengths
\citep[see][for a review]{nk11}. These pulsars could represent transition
objects that share similar observational properties with magnetars and radio
pulsars, providing the key to understanding the connection between the two
classes.

\object{\psr} is a young high-$B$ radio pulsar discovered in the Parkes
Multibeam Pulsar Survey \citep{ckl+00}. It has a spin period $P=408$\,ms and a
large period derivative $\dot P=4.0\times10^{-12}$, which together imply a
surface dipole $B$-field of $4.1\times10^{13}$\,G. It is one of the very few
pulsars with a measured braking index.\footnote{The braking index is defined as
$n\equiv\nu\ddot\nu/\dot\nu^2$, where $\nu$ is the spin frequency and
$\dot\nu$ and $\ddot\nu$ are its first and second time derivatives,
respectively.} \citet{wje11} reported $n=2.684\pm0.002$ using over 12 years
of radio timing data, which give a characteristic age of $P/[(n-1)\dot P]
=1.9$\,kyr. The true age could even be younger if the pulsar were born with a
spin period close to the present-day value and $n$ is constant. The pulsar is
associated with a 17\arcmin-diameter radio supernova remnant (SNR) shell
\object{\snr} \citep{cgk+01}. Based on H{\sc i} absorption measurements of the
SNR, \citet{cmc04} reported a source distance of $8.4\pm0.4$\,kpc.

\begin{deluxetable}{lccc}
\tablecaption{X-Ray Observations of \psr\ Used in This Study \label{table:obs}}
\tablewidth{0pt}
\tablehead{\colhead{ObsID}& \colhead{Date} & \colhead{Instrument / Mode} &
\colhead{Net Exp.\tablenotemark{a} (ks)}}
\startdata
\multicolumn{4}{c}{\xmm}\\\hline
\dataset[ADS/Sa.XMM#obs/0150790101]{0150790101} & 2003 Jun 26 & PN / Large
Window & 41.8 \\
& & MOS1 / Full Frame & 44.3 \\
& & MOS2 / Full Frame & 44.8 \\
\dataset[ADS/Sa.XMM#obs/0672790101]{0672790101} & 2011 Jun 14 & PN / Large
Window & 27.7 \\
& & MOS1 / Full Frame & 34.1 \\
& & MOS2 / Full Frame & 34.3 \\
\dataset[ADS/Sa.XMM#obs/0672790201]{0672790201} & 2011 Jun 30 & PN / Large
Window & 27.9 \\
& & MOS1 / Full Frame & 33.8 \\
& & MOS2 / Full Frame & 33.7 \\\hline
\multicolumn{4}{c}{\cxo}\\\hline
\dataset[ADS/Sa.CXO#obs/2833]{2833} & 2002 Mar 31 & ACIS-S / TE & 56.8 \\
\dataset[ADS/Sa.CXO#obs/4676]{4676} & 2004 Oct 31 & ACIS-S / TE & 60.5 \\
\dataset[ADS/Sa.CXO#obs/6153]{6153} & 2004 Nov 02 & ACIS-S / TE & 18.9
\enddata
\tablenotetext{a}{After removing periods of high background.}
\end{deluxetable}

In the X-ray band, the pulsar and SNR were detected with \emph{ASCA},
\emph{ROSAT}, the \emph{Chandra X-ray Observatory}, and \emph{XMM-Newton}
\citep{pkc+01,gs03,gs05,gkc+05,sk08}. The pulsar emission consists of thermal
and non-thermal components \citep{gkc+05,sk08} and shows strong pulsations
in the soft X-ray band \citep{gkc+05}. There is a faint, jet-like pulsar wind
nebula (PWN) extending 7\arcsec\ from the pulsar \citep{gs03,sk08}, but 
no radio counterpart has been found \citep{cgk+01}. The interior of SNR \snr\
shows faint, diffuse X-ray emission with spectral lines that indicate a
thermal origin \citep{gs05}. Recently, the \emph{Fermi Large Area Telescope}
detected $\gamma$-ray pulsations from \psr, making it the highest $B$-field
$\gamma$-ray pulsar yet known \citep{pkd+11}. Here we report on a detailed
study of \psr\ and SNR \snr\ using new \emph{XMM-Newton} observations together
with archival \xmm\ and \cxo\ data.

\section{Observations and Data Reduction}

New observations of \psr\ were carried out on 2011 June 14 and 30 with 
\emph{XMM-Newton}. The PN camera was operated in large window mode, which
has a time resolution of 48\,ms, while the MOS cameras were in full-frame
mode, with 2.6\,s time resolution, both with the medium filter. We also
reprocessed archival \xmm\ and \cxo\ data taken in 2002--2004. The \cxo\
observations were all made with the ACIS-S detector, which has 3.2\,s frame
time. Table~\ref{table:obs} lists the observation parameters of all data sets
used in this study. 

\xmm\ data reduction was performed using SAS 11.0 with the latest calibration
files. We processed the Observation Data Files with the tools
\texttt{emproc/epproc}, then employed \texttt{espfilt} to remove periods of
high background. The resulting exposure for the PN and MOS cameras are 97 and
112\,ks, respectively. Throughout the analysis below, we used events having
\texttt{PATTERN $\leq$ 12}, unless specified otherwise. \cxo\ data reduction
was carried out with CIAO 4.4 and CALDB 4.4.7. We applied the script
\texttt{chandra\_repro} to generate new \texttt{level=2} event files with the
latest time-dependent gain, charge transfer inefficiency correction and
sub-pixel adjustments. We obtained a net exposure of 136\,ks after rejecting
periods with background flares.

\section{Analysis and Results}
\subsection{Imaging}
\begin{figure*}[ht]
\epsscale{1.0}
\plotone{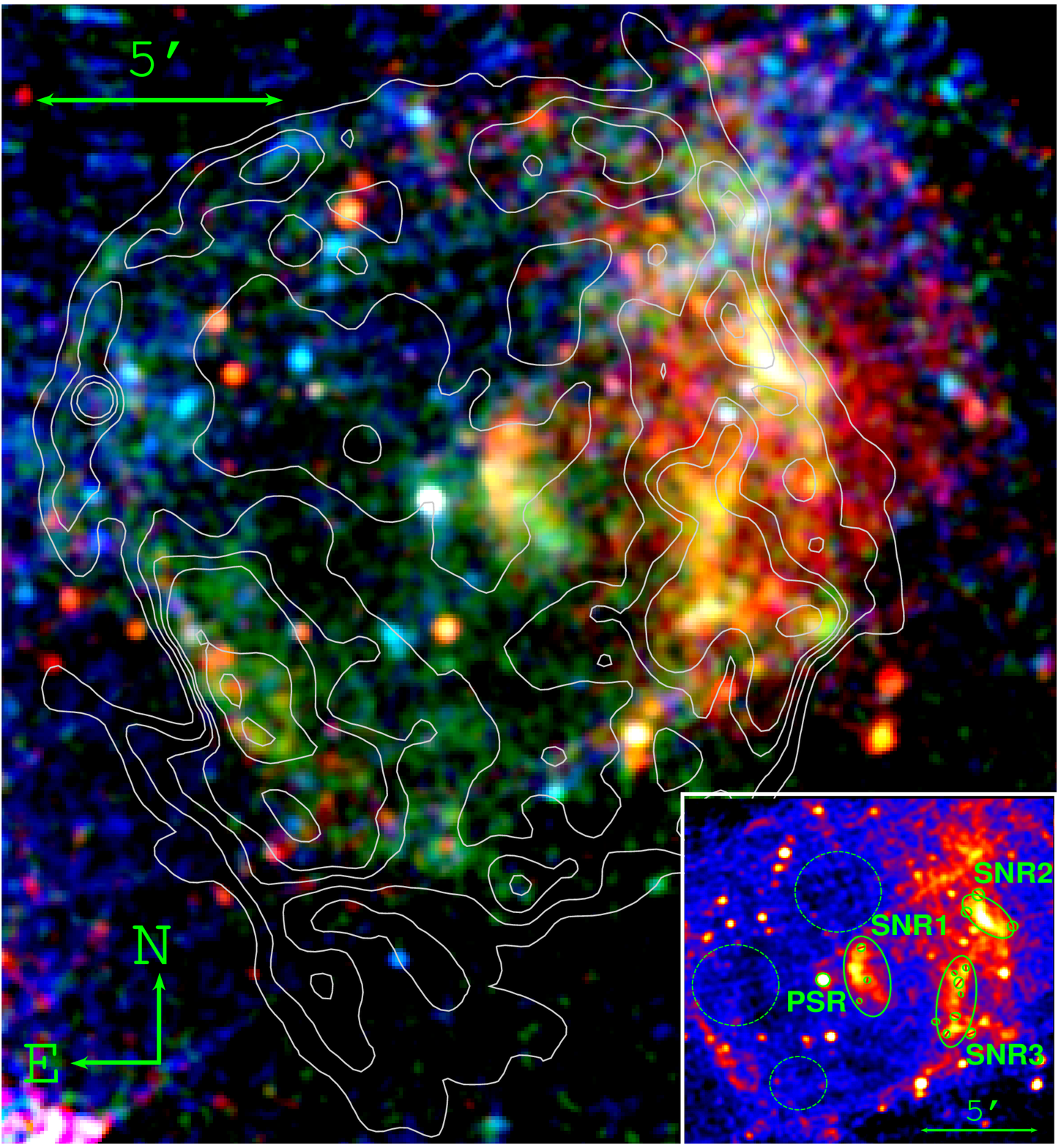}
\caption{Three-color image of \psr\ and \snr\ in the 0.3--1.5\,keV (red),
1.5--3\,keV (green), and 3--8\,keV (blue) bands, made with all
\xmm\ PN and MOS data listed in Table~\ref{table:obs} with a total exposure
of 100\,ks. The image is exposure-corrected, smoothed to 13\arcsec\ resolution,
and overlaid with 1.4-GHz radio contours from \citet{cgk+01}. Inset:
intensity map in the 0.3--8\,keV band, indicating the source and background
extraction regions for the spectral analysis. \label{fig:img}}
\end{figure*}

We generated individual exposure-corrected images for each PN and MOS
observation with the tasks \texttt{evselect} and \texttt{eexpmap}, then
combined them using \texttt{emosaic}. Figure~\ref{fig:img} shows a three-color
\xmm\ image overlaid with 1.4\,GHz radio contours from \citet{cgk+01} and a
broadband intensity image in the 0.3--8\,keV energy range. We have also
generated images with the \cxo\ data, but they are not shown here due to the
poorer sensitivity and sky coverage. As shown in the figure, \psr\ and a
partial shell of SNR \snr\ are detected. The SNR emission is clumpy and it
generally traces the brightest part of the 17\arcmin\,diameter radio shell,
except for the diffuse emission near the pulsar, which shows no radio
counterpart. The western half of the shell is brighter and redder than the
eastern half. Note that the southwest rim of the shell falls outside the PN
camera field of view, hence, the sensitivity is largely reduced.

\subsection{Timing}
The timing analysis was carried out with the PN data only, since only they
have sufficiently high-time resolution. We extracted source events from a
circular aperture of 18\arcsec\ radius centered on the pulsar. This radius was
determined by the tool \texttt{eregionanalyse} to yield the optimal
signal-to-noise ratio. We followed \citet{gkc+05} to correct for a timing
anomaly in the 2003 data; we did not find any other such anomaly in the other
observations. We divided the data into low-energy (0.5--2.5\,keV) and
high-energy (2.5--8\,keV) bands, and obtained $889\pm30$ and $414\pm20$ total
counts, respectively, with $\sim110$ and 120 background counts, respectively,
as estimated from nearby regions. The event arrival times were
first corrected to the solar system barycenter, then folded according to the
radio ephemerides. We used the published pulsar ephemeris from \citet{wje11}
for the 2003 data, while the ephemeris in 2011 was obtained from
contemporaneous radio observations with the Parkes radio telescope, as part of
the timing program for \emph{Fermi} \citep{wjm+10}.

\begin{figure}[ht]
\epsscale{1.2}
\plotone{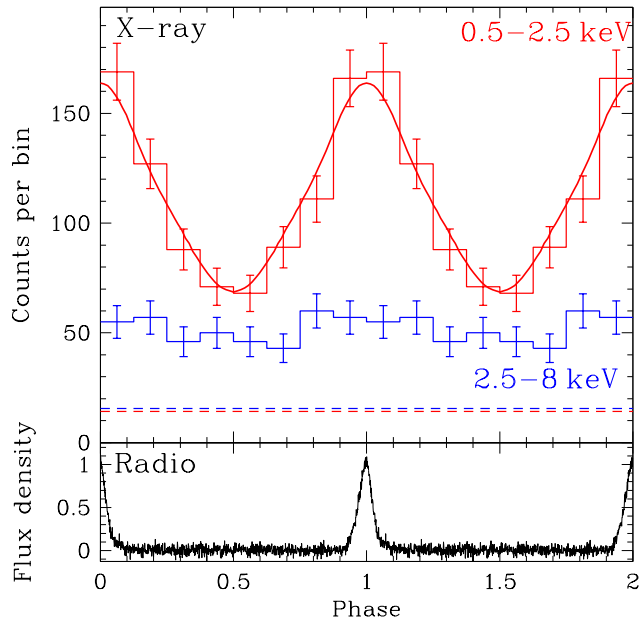}
\caption{X-ray pulse profiles of \psr\ from \xmm\ PN observations
in 0.5--2.5\,keV (red) and 2.5--8\,keV (blue). The solid
line shows the model lightcurve of the soft X-ray emission from a single hot
spot with a magnetized atmosphere (see Section~\ref{sect:psrthermal}). Background
levels are indicated by the dashed lines. Phase 0 corresponds to the radio
pulse at 1.4\,GHz shown in the lower panel. The radio flux density is in an
arbitrary unit. \label{fig:pulse}}
\end{figure}

First, we verified that the pulse profiles in 2003 and 2011 were consistent.
They were subsequently co-added in the analysis to boost the signal-to-noise
ratio. The resulting profile is shown in Figure~\ref{fig:pulse}. The pulsar
emission below 2.5\,keV is highly modulated. It exhibits a single-peaked
profile with good alignment with the radio pulse. A sinusoid provides an
adequate fit to the profile and it gives an rms pulsed fraction (PF) of
$0.38\pm0.03$. To compare with previous work, we follow \citet{gkc+05} and
calculate the ``max--min PF'' using PF=$(F_{\rm max}-F_{\rm min})/(F_{\rm
max}+F_{\rm min})$, where $F_{\rm max}$ and $F_{\rm min}$ are the maximum and
minimum background-subtracted counts in the binned profile, respectively. In
this way, we obtain max--min PF of $0.48\pm0.12$ in the 0.5--2.5\,keV range.
This is somewhat lower than the value $0.74\pm0.14$ reported by
\citet{gkc+05}, and the discrepancy could be attributed to Poisson
fluctuations or to difference in phase binning. On the other hand, no
pulsations are detected above 2.5\,keV. This is consistent with the findings
of \citet{gkc+05} and with the \emph{RXTE} results in the 2--60\,keV range
reported by \citet{pkd+11}. We place put an upper limit of PF $<10\%$ in the
2.5--8\,keV range. 

\subsection{Spectroscopy}
We carried out the spectral analysis with 
XSPEC\footnote{\url{http://heasarc.gsfc.nasa.gov/xanadu/xspec/}} 12.7.1. Spectra
were grouped to at least 20 counts per bin, and only single and double events
(i.e.\ \texttt{PATTERN $\leq$ 4}) were used for the PN data to ensure the best
spectral calibration. We restricted the fits in the 0.5--8\,keV. A
photoelectric absorption model \texttt{tbabs} by \citet{wam00} was employed
throughout our study, and we adopted the abundances given by the same authors.

\subsubsection{Phase-averaged Pulsar Spectroscopy}
Spectra of \psr\ were extracted from 18\arcsec-radius apertures from the \xmm\
and \cxo\ data. Background spectra were extracted from nearby regions free
from SNR emission shown in Figure~\ref{fig:img}. We have also tried using
annular background regions centered on the pulsar, and the results are very
similar. There are $1062\pm37$, $500\pm25$, $442\pm24$, and $908\pm35$ net
pulsar counts in 0.5--8\,keV from the PN, MOS1, MOS2, and ACIS detectors,
respectively. A comparison of different spectra shows no variability between
epochs. Therefore, we performed a joint fit to all the data. Given the low
number of source counts, we did not attempt to account for the
cross-calibration uncertainty between instruments \citep[see][]{tgp+11}. 

\begin{figure}[th]
\includegraphics[width=0.48\textwidth,clip=true]{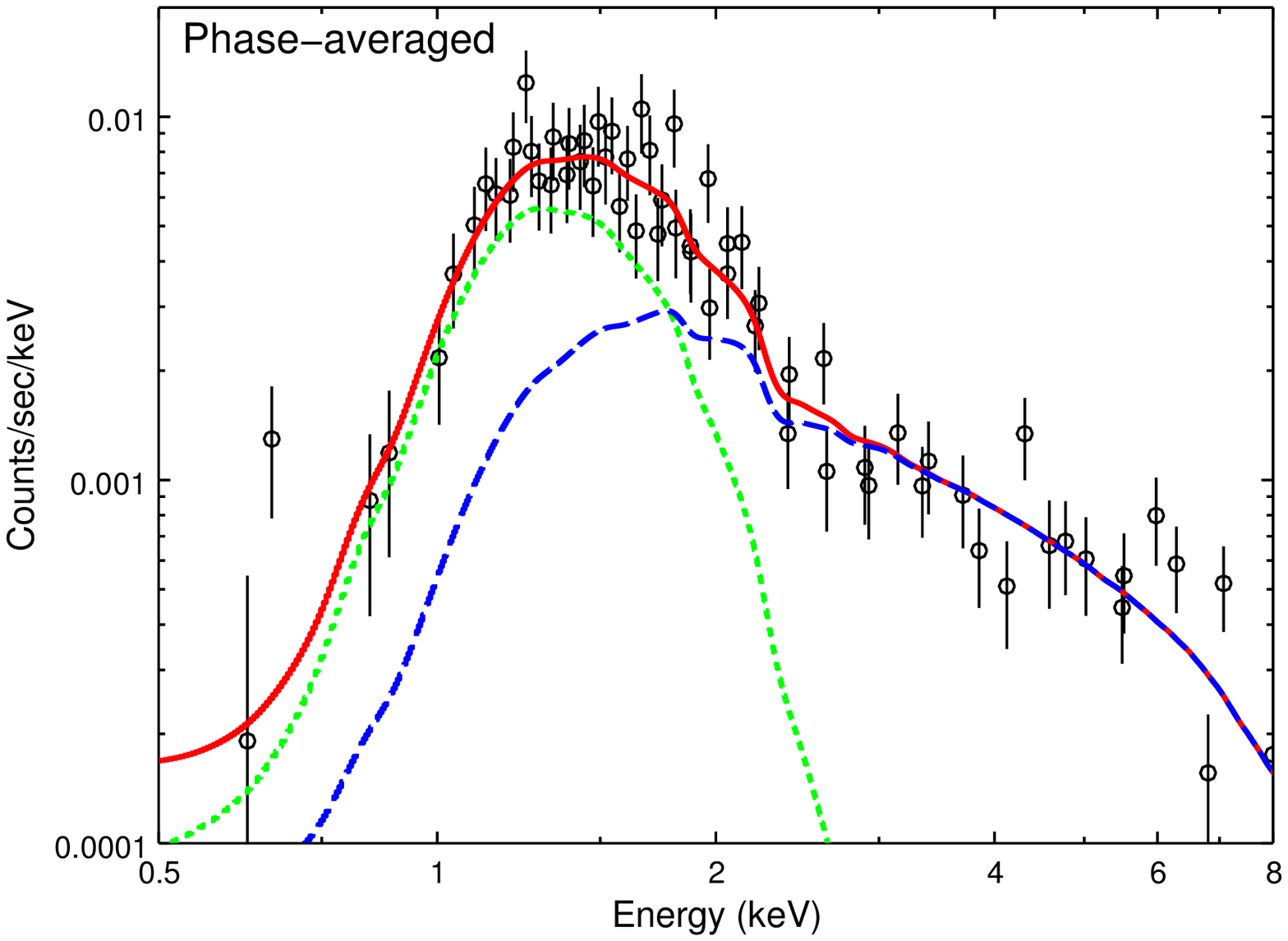}
\includegraphics[width=0.48\textwidth,clip=true]{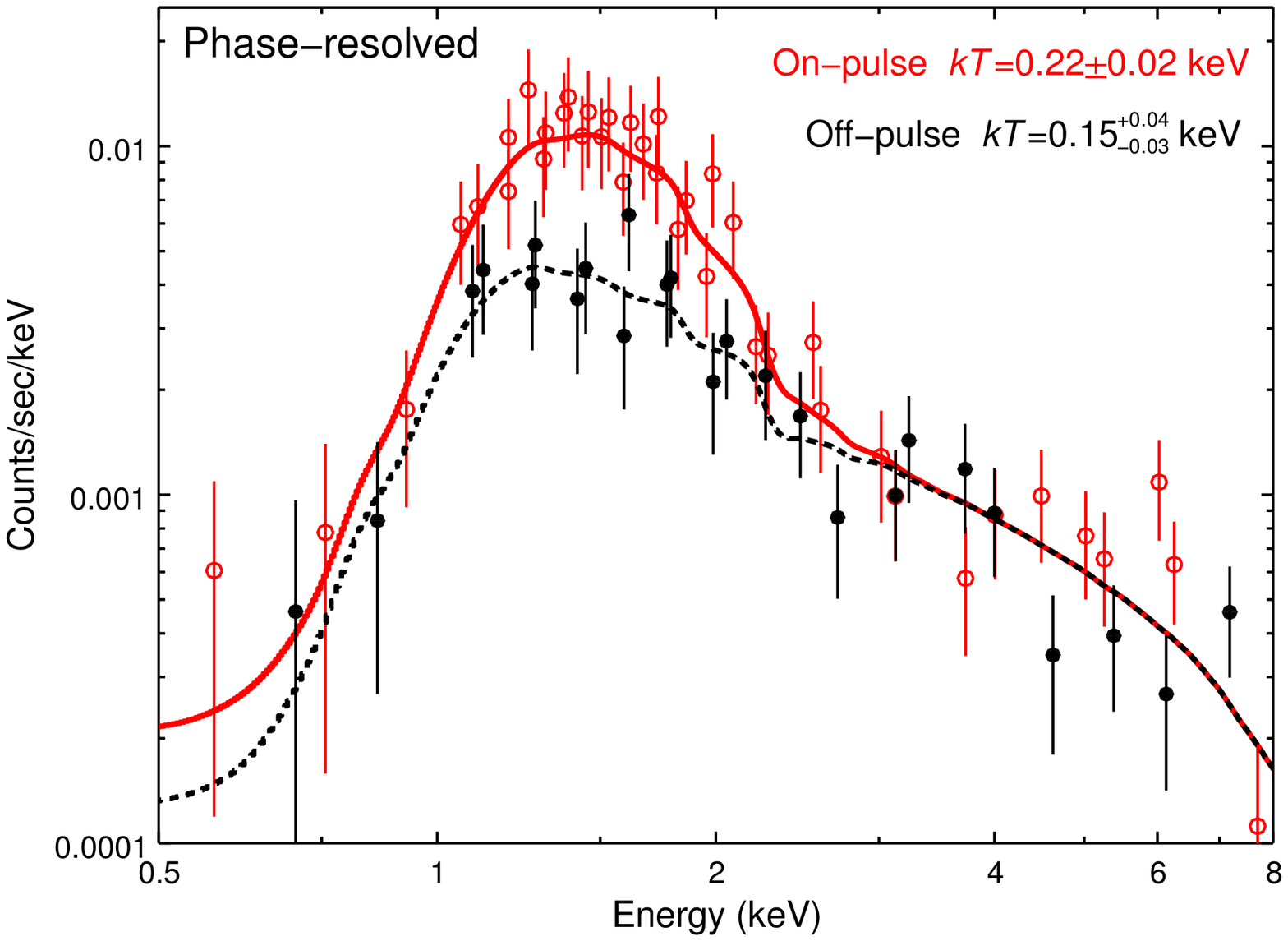}
\caption{Top: best-fit blackbody plus power-law model for the
phase-averaged spectrum of \psr. The thermal component is shown by the dotted
green line and the non-thermal component, which composes of pulsar and PWN
emission, is shown by the dashed blue line. Only the PN spectra are shown for
clarity. Bottom: on-pulse (red open circle) and off-pulse (black filled
circle) PN spectra of the pulsar, fitted with the blackbody plus power-law
model. See Section~\ref{sect:phspec} for details. \label{fig:psrspec}}
\end{figure}

To account for the PWN contamination, we employed a PWN component in the
spectral model, which was extracted from the \cxo\ data using an annular
aperture of 2.5\arcsec--18\arcsec. We fitted the PWN spectrum using an absorbed
power-law (PL) model with the column density (\nh) fixed at the pulsar value.
The PWN component was then held fixed during the fit of the pulsar spectrum.
Once a new value of \nh\ was determined, we re-fit the PWN spectrum and the
whole process was carried out iteratively until consistent spectral parameters
were found. We first tried simple models, including an absorbed blackbody
(BB) and an absorbed PL for the pulsar spectrum. However, they fail to
describe the data, and we found that an absorbed BB+PL model provides a much
better fit ($\chi^2_\nu=0.97$). This gives a high effective temperature of
$kT=0.21\pm0.04$\,keV with a emission radius of $3^{+4}_{-1}$\,km (for a
source distance of 8.4\,kpc), and a PL photon index of $2.1\pm0.8$. The BB
component has an absorbed flux of $1.2\pm0.3\times10^{-14}$\ergcm\ in
the 0.5--8\,keV energy range, and the non-thermal components have absorbed
fluxes of $2.3\pm0.4\times10^{-14}$\ergcm\ and $2.8\times10^{-14}$\ergcm\ from
the pulsar and the PWN, respectively. The best-fit parameters are listed in
Table~\ref{table:psrspec} and the PN spectrum is shown in
Figure~\ref{fig:psrspec}(a).

Following previous studies, we also tried fitting the pulsar spectra with a
neutron star atmosphere model \citep[NSA;][]{zps96} plus a PL model. The
former describes a fully ionized hydrogen atmosphere in a $B$-field of
$10^{13}$\,G. During the fits, the stellar mass was fixed at 1.4$M_\odot$ and
we assumed uniform thermal emission over the entire surface of 13\,km radius.
This yields a fit that is equally as good as the BB+PL model, but with a
significantly lower surface temperature of $0.08^{+0.03}_{-0.02}$\,keV. The
flux normalization suggests a distance of $2.4_{-1.8}^{+5.6}$\,kpc, which is
lower than the source distance of 8.4\,kpc. Table~\ref{table:psrspec}
summarizes the best-fit parameters of the BB+PL and NSA+PL fits. All
uncertainties quoted are at the 90\% confidence level. The parameters are
fully consistent with previous studies \citep{gkc+05,sk08}, and are better
constrained in our case.
For the case of magnetars, BB+PL spectra are often observed
\citep[e.g.,][]{nkd+11} and the PL component is interpreted as magnetospheric
upscattering of thermal photons from the surface emission \citep{tlk02}. To
check if this is the case for \psr, we fitted the pulsar spectra with a
resonant cyclotron scattering model \citep[RCS;][]{rzt+08}, but obtained very
poor results ($\chi^2_\nu\approx2$), mainly because the model provides too little
flux above 3\,keV to account for the observed hard X-ray tail.

\begin{deluxetable}{lcc}
\tablecaption{Best-fit Blackbody Plus Power-law (BB+PL) and Neutron Star
Atmosphere Plus Power-law (NSA+PL) Models to the Phase-averaged Spectrum of
\psr \label{table:psrspec}} \tablewidth{0pt}
\tablehead{\colhead{Parameter} & \colhead{BB+PL} &
\colhead{NSA\tablenotemark{a}+PL} }
\startdata
\nh\ ($10^{22}$\cms) & $2.2^{+0.5}_{-0.4}$ & $2.5^{+0.6}_{-0.5}$\\
$kT^\infty$ (keV) & $0.21\pm0.04$ & $0.08_{-0.02}^{+0.03}$\\
$R^\infty$ (km) & $3_{-1}^{+4}$ & 13\tablenotemark{b} \\
Distance (kpc) & 8.4\tablenotemark{b} & $2.4_{-1.8}^{+5.6}$ \\
$\Gamma_{\rm PSR}$ & $2.1\pm0.8$ & $2.1\pm0.8$ \\
$\Gamma_{\rm PWN}$ & 1.31\tablenotemark{b} & 1.46\tablenotemark{b} \\
$f_{\rm bb/nsa}^{\rm abs}$ ($10^{-14}$\ergcm) & $1.2\pm0.3$ &
$1.4_{-0.3}^{+0.2}$ \\
$f_{\rm bb/nsa}^{\rm unabs}$ ($10^{-14}$\ergcm) & $16_{-7}^{+16}$ &
$43_{-25}^{+97}$ \\
$f_{\rm pl}^{\rm abs}$ ($10^{-14}$\ergcm) & $2.3\pm0.4$ & $2.3\pm0.4$ \\
$f_{\rm pl}^{\rm unabs}$ ($10^{-14}$\ergcm) & $5_{-2}^{+6}$ & $5_{-2}^{+5}$ \\
$f_{\rm PWN}^{\rm abs}$ ($10^{-14}$\ergcm) & 2.8\tablenotemark{b} &
2.7\tablenotemark{b}\\
$f_{\rm PWN}^{\rm unabs}$ ($10^{-14}$\ergcm) & 4.0\tablenotemark{b} &
4.3\tablenotemark{b}\\
$\chi^2_\nu$/dof & 0.97/161 & 0.97/161
\enddata
\tablecomments{All uncertainties are 90\% confidence intervals (i.e.,
1.6$\sigma$). $f^{\rm abs}$ and $f^{\rm unabs}$ are the absorbed and
unabsorbed fluxes, respectively, in the 0.5--8\,keV energy range.} 
\tablenotetext{a}{-- A $B$-field strength of $10^{13}$\,G is assumed in the
NSA model.}
\tablenotetext{b}{-- Held fixed during the fit.}
\end{deluxetable}

\subsubsection{Phase-resolved Pulsar Spectroscopy}
\label{sect:phspec}
To understand the nature of the X-ray pulsations, we performed phase-resolved
spectroscopy using the PN data. Based on the pulse profile in
Figure~\ref{fig:pulse}, we extracted the off-pulse spectra from phase
0.25--0.75, and the on-pulse spectra from the rest of the phase range. There
are $673\pm29$ and $385\pm23$ net counts for the on- and off-pulse emission,
respectively, in the 0.5--8\,keV energy range. The spectra were grouped with
at least 20 counts per bin, and fitted to the BB+PL model in
the 0.5--8\,keV range. As shown in Figure~\ref{fig:psrspec}(a), the pulsar
emission above 2.5\,keV is dominated by non-thermal emission, which shows no
pulsations (see Figure~\ref{fig:pulse}). Therefore, the non-thermal components
(from both the pulsar and the PWN) were held fixed at the phase-averaged
values during the fits. We also fixed the absorption column density at the
best-fit phase-averaged value of $N_{\rm H}=2.2\times10^{22}$\,cm$^{-2}$. We
note that the non-thermal flux is not enough to account for the off-pulse
emission, suggesting the presence of thermal emission at pulse minimum. We
found a higher BB temperature of $kT=0.22\pm0.02$\,keV with an
effective radius of $R=3.0^{+1.0}_{-0.8}$\,km for the on-pulse emission. The
off-pulse emission shows a hint of a lower temperature of
$kT=0.15^{+0.04}_{-0.03}$\,keV and a larger radius of $7_{-7}^{+13}$\,km. The
best-fit spectral models are shown in Figure~\ref{fig:psrspec}(b). We have
also tried fitting the difference spectrum between on- and off-pulse emission.
While it is consistent with a BB model, there are large uncertainties
in the spectral parameters due to poor data quality.

\begin{deluxetable*}{lccccccccc}
\tablecaption{Best-fit PSHOCK Plus Power-law (PS+PL) and Two-temperature
PSHOCK (PS+PS) Models for the Spectra of SNR \snr \label{table:snrspec}}
\tabletypesize{\small}
\tablewidth{0pt}
\tablehead{& \multicolumn{2}{c}{Region 1} && \multicolumn{2}{c}{Region 2} &&\multicolumn{2}{c}{Region 3}\\
\cline{2-3} \cline{5-6} \cline{8-9}
\colhead{Parameter} & \colhead{PS+PL}& \colhead{PS+PS} && \colhead{PS+PL}&
\colhead{PS+PS} && \colhead{PS+PL}& \colhead{PS+PS}
}
\startdata
\nh\ ($10^{22}$\cms) & $2.17^{+0.07}_{-0.08}$ & $2.17^{+0.10}_{-0.08}$ &&
$1.9_{-0.2}^{+0.5}$ & $0.9\pm0.2$ && $1.15^{+0.07}_{-0.03}$ & $1.0\pm0.1$\\
$kT_1$ (keV) & $0.18\pm0.02$ & $0.18^{+0.3}_{-0.2}$ && $0.11^{+0.05}_{-0.02}$ &
$0.74^{+0.11}_{-0.03}$ && $0.66^{+0.03}_{-0.02}$ & $0.66\pm0.03$ \\
$\tau_1$ ($10^{10}$\,s\,cm$^{-3}$) & $40_{-20}^{+40}$ & $90_{-40}^{+90}$ &&
 $>150$ & $>30$ && $>200$ & $>100$ \\
EM$_1$ ($10^{56}$\,cm$^{-3}$) & $360_{-140}^{+200}$ & $330^{+160}_{-120}$ &&
$1100^{+3000}_{-400}$ & $0.5\pm0.2$ && $1.9^{+0.4}_{-0.5}$ & $1.5^{+0.6}_{-0.4}$ \\
$\Gamma$ & $2.9\pm0.1$ & \nodata && $3.5^{+0.2}_{-0.1}$ & \nodata && $3.5\pm0.2$ &
\nodata \\
$kT_2$ (keV) & \nodata & $1.8^{+0.2}_{-0.1}$ && \nodata & $1.8\pm0.2$ &&
 \nodata & $1.0\pm0.1$ \\
$\tau_2$ ($10^{10}$\,s\,cm$^{-3}$) & \nodata & $<0.2$ &&
 \nodata & $<0.08$ && \nodata & $<0.08$ \\
EM$_2$ ($10^{56}$\,cm$^{-3}$) & \nodata & $4.2^{+0.4}_{-0.6}$ && \nodata &
$2.2\pm0.3$ && \nodata & $4.2^{+0.8}_{-0.7}$ \\
$f_1^{\rm abs}$ & $0.64\pm0.03$ & $0.45\pm0.02$ &&
 $0.15^{+0.04}_{-0.03}$ & $0.22^{+0.07}_{-0.06}$ &&
 $0.51\pm0.08$ & $0.48^{+0.09}_{-0.08}$ \\
$f_1^{\rm unabs}$ & $330^{+100}_{-80}$ & $300^{+110}_{-80}$ &&
 $330^{+700}_{-150}$ & $1.1^{+0.8}_{-0.5}$ &&
 $3.5^{+0.9}_{-0.8}$ & $2.9^{+1.0}_{-0.8}$ \\
$f_2^{\rm abs}$ & $1.46\pm0.06$ & $1.3^{+0.07}_{-0.05}$ &&
 $1.5\pm0.05$ & $0.88\pm0.08$ &&
 $0.73_{-0.09}^{+0.10}$ & $0.72^{+0.08}_{-0.09}$ \\
$f_2^{\rm unabs}$ & $6^{+0.6}_{-0.5}$ & $7^{+7}_{-4}$ &&
 $6^{+2}_{-1}$ & $1.8^{+0.9}_{-0.1}$ &&
 $3.7\pm0.7$ & $2.1^{+1.8}_{-0.2}$ \\
$\chi^2_\nu$/dof & 1.21/1175 & 1.21/1174 && 1.19/363 & 1.18/362 && 1.35/782
& 1.31/781
\enddata
\tablecomments{The subscripts 1 and 2 correspond to the soft and hard 
components, respectively. All uncertainties are 90\% confidence intervals
(i.e.\ 1.6$\sigma$). The volume emission measure is given by $\mathrm{EM}=\int
n_e n_{\rm H}\,{\rm d}V$, where $n_e$ and $n_{\rm H}$ are shocked electron
and hydrogen number densities, respectively. $\tau=n_et$ is the ionization
timescale. The absorbed and unabsorbed fluxes ($f^{\rm abs}$ and $f^{\rm
unabs}$) are estimated from the \xmm\ data, in units of $10^{-13}$\ergcm\ in
the 0.5--8\,keV energy range.}
\end{deluxetable*}

\subsubsection{SNR Spectroscopy}
We extracted spectra of SNR \snr\ from the three brightest regions (see
Figure~\ref{fig:img}) from the \xmm\ and \cxo\ data. Our region 1 roughly
corresponds to the ``east'' region in \citet{gs05}, and regions 2 and 3
correspond to their ``west'' region. Point sources inside these regions were
identified and excluded based on the exposure-corrected \cxo\ and \xmm\
images. We note that region 2 is not covered by any \cxo\ observations,
therefore, only the \xmm\ spectra are used in this case. Similarly, only one
\cxo\ exposure is useful for region 3. Since the SNR covers a large part
of the \cxo\ and \xmm\ PN field of view, we estimated the background spectra
using the blank-sky data provided by the calibration teams \citep{cr07}, which
were taken from nearby regions in the Galactic plane. We also tried extracting
the background spectra from source-free regions far off-axis in our
observations and obtained similar results, although this way the sky background
tends to be underestimated due to vignetting. After background subtraction,
we obtained $1.1\times10^4$ \xmm\ counts and 5300 \cxo\ counts for region 1,
4900 \xmm\ counts for region 2, and 7900 \xmm\ counts and 2000 \cxo\ counts
for region 3 in 0.5--7.5\,keV range.

\begin{figure}[th]
\includegraphics[width=0.48\textwidth,clip=true]{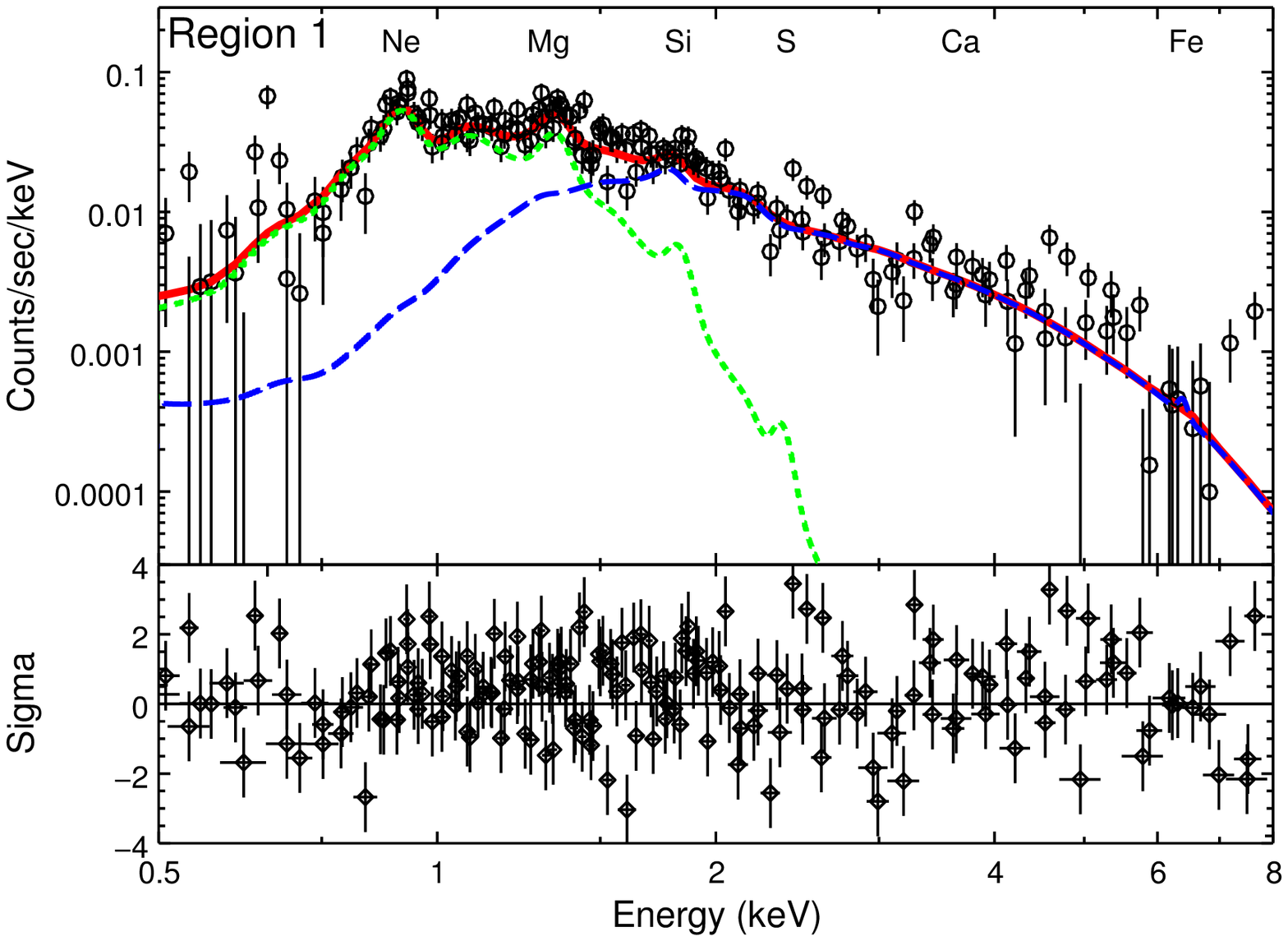}
\includegraphics[width=0.48\textwidth,clip=true]{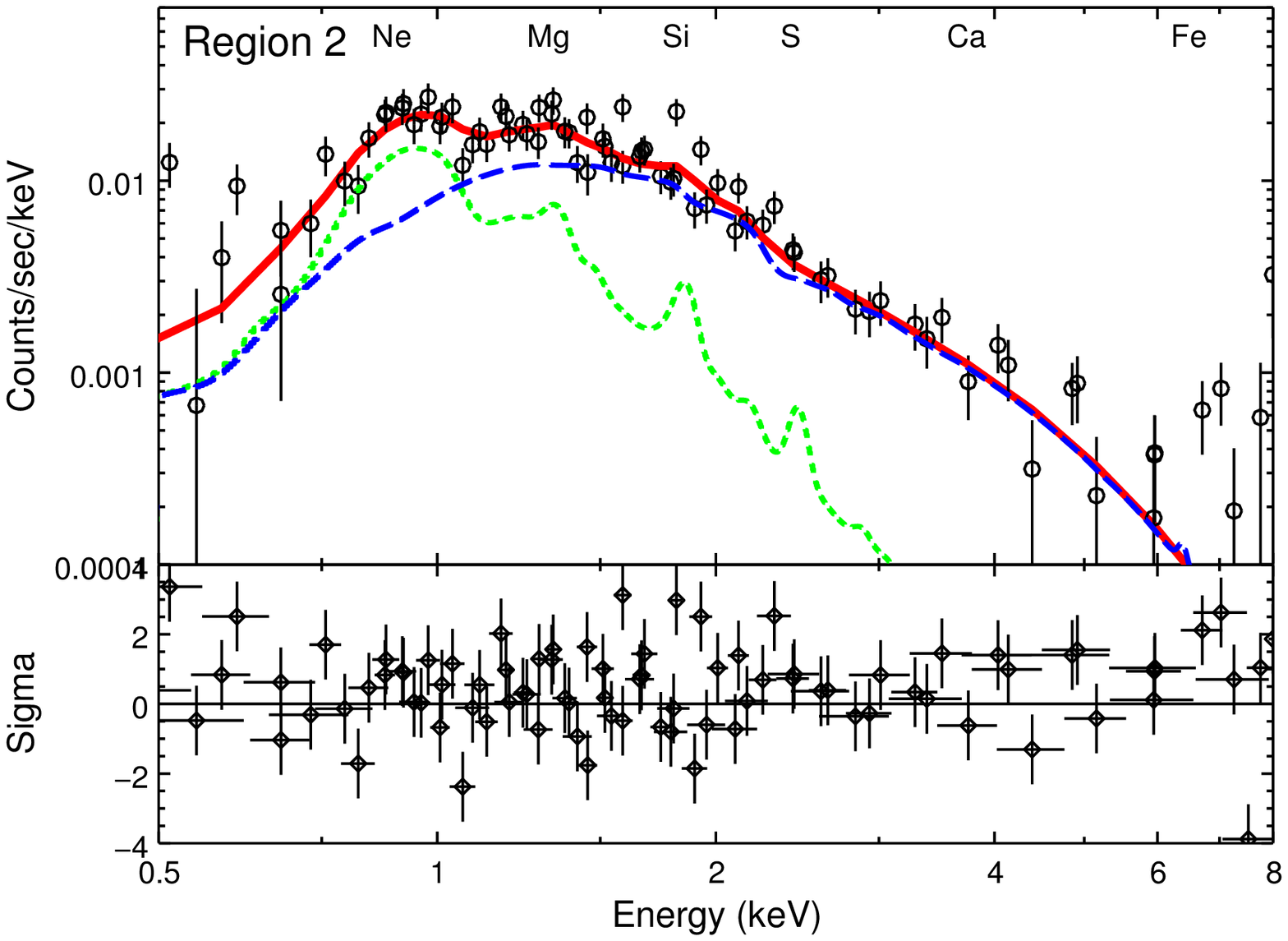}
\includegraphics[width=0.48\textwidth,clip=true]{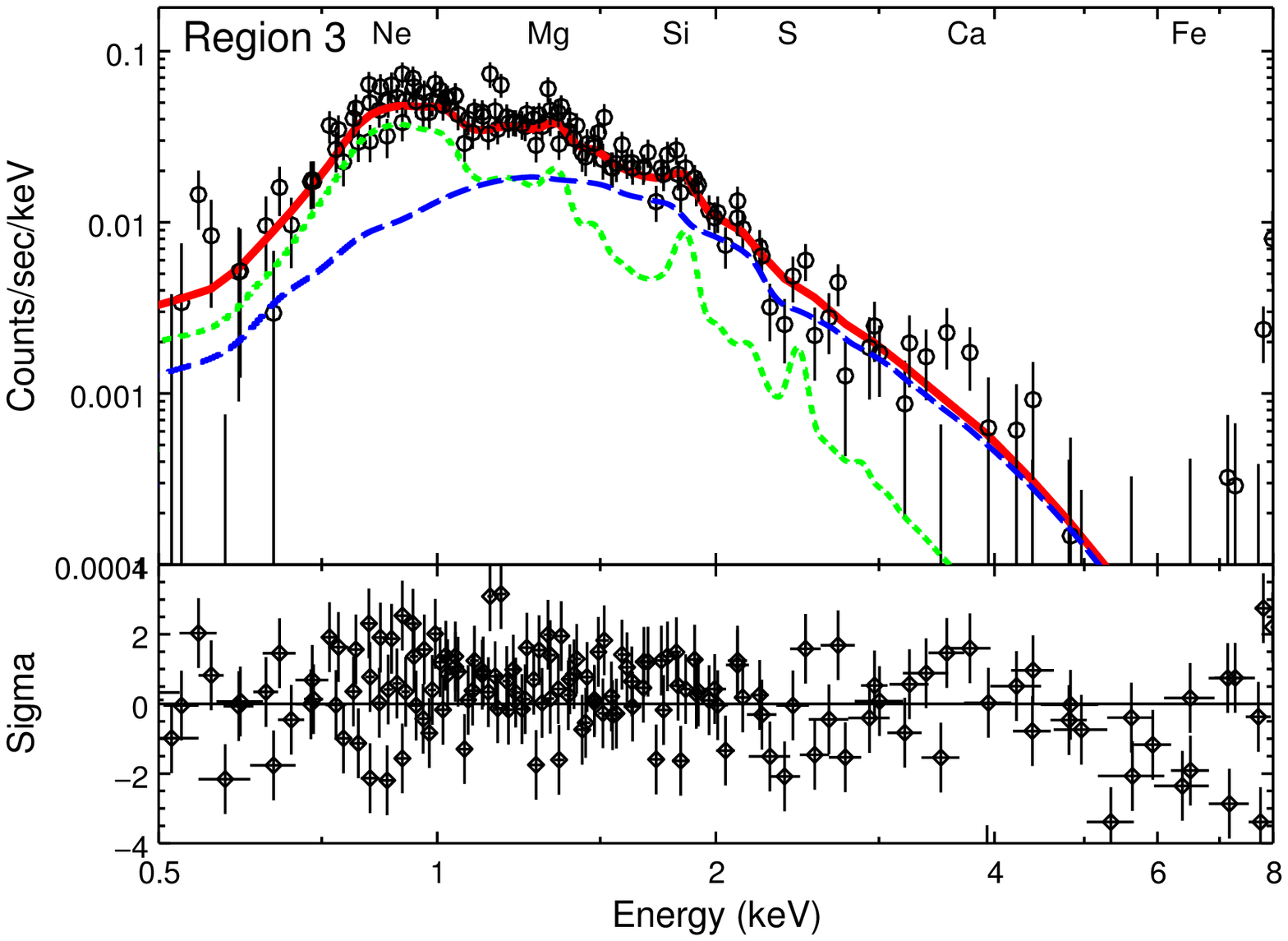}
\caption{Best-fit two-temperature PSHOCK models to the spectra of SNR \snr.
The high- and low-temperature components are indicated by the dashed blue
lines and dotted green lines, respectively. For clarity, only the PN spectra
are shown and all spectra are grouped with a minimum of 50 counts per bin.
\label{fig:snrspec}}
\end{figure}

Emission lines are found in all the spectra, albeit rather weak, indicating a
thermal origin for the emission. For each region, we performed a joint fit to
the spectra with all parameters linked, and we introduced a constant scaling
factor between the \xmm\ and \cxo\ spectra to account for the
cross-calibration uncertainty \citep{tgp+11}. We first tried the spectral
parameters reported by \citet{gs05}, but these gave very poor fits. Using a
non-equilibrium ionization thermal model \citep[PSHOCK;][]{blr01} with solar
abundances, we obtained reasonable results ($\chi^2_\nu=$1.3--1.5). The fits
can be further improved by allowing the abundances to vary (with the VPSHOCK
model) or by adding a hard component, and the improvements are statistically
significant as indicated by $F$-tests. However, since the emission lines are
weak, the VPSHOCK fits require rather low abundances (0.1--0.5) for all
elements between Ne and Fe. Therefore, we believe that a two-component model
could be a more physical description of the thermal emission. In this case,
the hard component can be described by either a non-thermal PL with
photon indices $\Gamma=2.9$--3.5 or by a high temperature ($kT=1$--1.8\,keV)
PSHOCK model with very low ionization timescales ($\tau \lesssim
10^9$\,s\,cm$^{-3}$) and solar abundances, and they provide the best fits
overall ($\chi^2_\nu\approx$1.2). The results are listed in
Table~\ref{table:snrspec} and the best-fit two-temperature models are plotted
in Figure~\ref{fig:snrspec}. In all the fits, the soft component has a large
ionization timescales ($\tau \gtrsim10^{12}$\,s\,cm$^{-3}$). The \cxo\ flux
is $\sim30$\% higher than the \xmm\ flux, which is not atypical
\citep[see][]{tgp+11}.

\section{Discussion}
In the following discussion, we adopt a source distance of 8.4\,kpc and an age
of 1.9\,kyr. Although the true age of the system could be different than the
pulsar's characteristic age, we stress that our discussion should remain
qualitatively valid for a system age younger than a few thousand years.

\subsection{\psr}
\subsubsection{Thermal Emission}
\label{sect:psrthermal}
\psr\ is among the youngest radio pulsars with thermal emission detected. As
indicated in Figure~\ref{fig:psrspec}(a), the pulsar flux below 2.5\,keV is
mostly thermal, and it is highly pulsed and shows good phase alignment
with the radio pulse (Figure~\ref{fig:pulse}). Combining these two results
suggests that the thermal emission could be dominated by a single hot spot
near the magnetic pole. Further support for this picture is provided by the
phase-resolved spectroscopy, which suggests a hint of a higher temperature for
the on-pulse emission than the off-pulse emission (Figure~\ref{fig:psrspec}(b)).
The BB+PL fit to the phase-averaged spectrum gives a high temperature of
$0.21\pm0.04$\,keV ($=2.4\pm0.5$\,MK) with an emission radius of
$3^{+4}_{-1}$\,km, implying a bolometric luminosity of $1.9^{+1.9}_{-0.8}
\times10^{33}$\ergs\ for the thermal emission. The BB temperature and
luminosity are among the highest of radio pulsars, even when compared to
objects with similar age \citep[see][]{okl+10,nk11,zkm+11}. Since the stellar
surface temperature is non-uniform, the NSA fits only give the average
temperature and it cannot be directly compared to the BB temperature. Indeed,
the bolometric luminosity above is equivalent to that of a 13\,km radius star
with a uniform temperature of 0.096\,keV (=1.1\,MK), comparable to the
best-fit NSA temperature of $0.08^{+0.03}_{-0.02}$\,keV.

\citet{gkc+05} and \citet{sk08} pointed out that the BB luminosity and radius
are too large to be reconciled with polar cap heating by return currents from
the magnetosphere. Hence, the thermal emission is likely from cooling and its
unusual properties, e.g.\ high temperature and large modulation, could be
related to the strong magnetic field of \psr.
\citet{plm+07} noticed an apparent correlation between the BB temperature $T$
and the dipole $B$-field $B_d$ of neutron stars, with $T\propto \sqrt{B_d}$
spanning over three orders of magnitude in $B_d$, from radio pulsars to magnetars.
They attributed this to crustal heating by magnetic field decay, and
subsequently studied the ``magneto-thermal evolution'' of neutron stars
\citep{apm08a,apm08b,pmg09}. In their model, heating from field decay
increases the magnetic diffusivity and thermal conductivity of the crust,
which in turn accelerates the field decay. As a result, neutron stars born
with initial $B$-fields $>5\times 10^{13}$\,G could have a surface
temperature well above the minimal cooling scenario \citep{apm08a,kpy+09}.

The thermal emission of \psr\ offers an interesting test case for the
above theory. While the BB temperature we obtained is comparable to the
predicted value at the pole \citep{apm08a}, the bolometric luminosity of $\sim
10^{33}-10^{34}$\ergs\ could be easily explained by passive cooling
\citep[e.g.][]{plp+04}, thus, providing no evidence for energy injection from
$B$-field decay. Indeed, the higher temperature at the pole could be
attributed to anisotropic heat conduction. Since free electrons, which are the
main heat carriers, are confined to the magnetic field lines, heat transport
perpendicular to the field is strongly suppressed \citep{gh83}. This acts as a
``heat blanket'' at the magnetic equator, while the magnetic poles are in
thermal equilibrium with the core. As a result, the surface temperature
distribution is non-uniform with warmer spots at the magnetic poles separated
by cooler regions at lower magnetic latitude \citep[e.g.][]{gkp04,sl12}.

It is believed that the internal magnetic field of a neutron star likely
consists of a toroidal component, which could be generated by differential
rotation that wraps the poloidal field in a proto-neutron star
\citep[e.g.][]{td93}. \citet{gkp06} showed that the presence of a toroidal
$B$-field in the crust could further suppress the heat transport from the core
towards the magnetar equator, because that requires crossing the toroidal
field lines. As a result, this would enhance the heat blanket effect.
Moreover, since the toroidal field component is symmetric about the magnetic
equator but the poloidal component is antisymmetric, the total $B$-field would
be asymmetric. Hence, the thermal emission from one magnetic pole could
dominate over another, which could help explain the single-peaked pulse
profile we found. Furthermore, the PF could be boosted by effects of
limb-darkening and magnetic beaming in the neutron star atmosphere
\citep[e.g.,][]{psv+94,rrm97,vl06}. Magnetic beaming is due to anisotropic
scattering and absorption of photons in magnetized plasmas. The resulting
radiation from a hot spot consists of a narrow pencil beam $<5\arcdeg$ along
the magnetic field and a broad fan beam at intermediate angles of
20\arcdeg--60\arcdeg\ \citep{zps+95,vl06}.

To test these ideas, we employed a magnetized partially ionized hydrogen
atmosphere model \citep{hpc08}, which accounts for the magnetic and
relativistic effects, to generate X-ray spectra and pulse profiles with a
similar procedure as \citet{ho07}. We took a surface $B$-field of
$4\times10^{13}$\,G and surface redshift of $1+z=1.19$, and adopted the
viewing geometry of magnetic inclination angle $\alpha\approx 125\arcdeg$ and
spin-axis inclination angle $\zeta\approx145 \arcdeg$ obtained from modeling
of the pulsar $\gamma$-ray lightcurve and radio polarization profile
\citep{pkd+11}. Vacuum polarization is ignored here, since it is not important
for the total surface emission at this field strength \citep{vl06}. We tried
different model parameters and found that two identical antipodal hot spots
give a very low PF of $\sim5$\%. The observed single-peaked pulse profile can
be produced if the temperature of one hot spot is higher than the other. Our
best-fit model is given by a hot spot with effective temperature (observed at
infinity) of $kT=0.13$\,keV (i.e.\ 1.5\,MK) and an area of one third of the
stellar surface, and for the other hot spot, we place a temperature upper
limit of 0.08\,keV assuming a similar area. This is shown by the solid line in
Figure~\ref{fig:pulse}. This temperature is lower than the best-fit BB
temperature for the on-pulse spectrum (see Figure~\ref{fig:psrspec}(b)) and also
the emitting area here is larger. These are typical characteristics of an
atmosphere model when compared to a simple BB, since non-gray opacities in the
former cause higher energy photons to be seen from deeper and hence hotter
layers in the atmosphere \citep[e.g.][]{rom87,szp+92}. Note that the model
suggests a rather large emitting area. This is because given the viewing
geometry, the observer's line of sight would cut through the fan beam twice
during one rotation. Therefore, a much smaller hot spot would produce a
double-peaked pulse profile. Finally, our model predicts a decreasing
``max--min PF'' with energy, from 0.5 at 0.5\,keV to 0.4 at 2\,keV, however,
this difference is too small to be detected given our data quality. Deeper
observations are needed to confirm this.

As a caveat, we note that the viewing geometry obtained from the $\gamma$-ray
lightcurve is model dependent. If the constraints on $\alpha$ and $\zeta$ are
relaxed, the high PF could also be explained by some extreme viewing
geometries, such that only one pole is observable \citep[see][]{gkc+05}.

\begin{deluxetable*}{lcccccccc}
\tablecaption{High-$B$ Radio Pulsars with Detected Pulsed Thermal Emission 
\label{table:highbpsr}} \tabletypesize{\small}
\tablewidth{0pt}
\tablehead{\colhead{Name} & \colhead{$P$} & \colhead{$B$} & \colhead{$\tau_c$}
& \colhead{Distance} & \colhead{$kT$} & \colhead{$L_{\rm bol}$} & \colhead{PF} &
\colhead{Reference}\\
& \colhead{(s)} & \colhead{($10^{13}$\,G)} & \colhead{(kyr)} & \colhead{(kpc)}
& \colhead{(keV)} & \colhead{($10^{32}$\ergs)} }
\startdata
J1119$-$6127 & 0.41 & 4.1 & 1.7 & 8.4& $0.21\pm0.04$ &
$19^{+19}_{-8}$ & $0.48\pm0.12$  & This work\\  
J1819$-$1458 & 4.3  & 5.0 & 117 & $\sim$3.6& $0.12\pm0.02$ & $1.2^{+2}_{-0.8}$
& $0.37\pm0.05$ & 1 \\
J1718$-$3718 & 3.4  & 7.4 & 34  & $\sim$4.5& $0.19\pm0.02$ &
$4^{+5}_{-2}$ & $0.60\pm0.13$ & 2\\
\enddata
\tablecomments{The characteristic ages ($\tau_c$) are inferred from the spin
parameters. $kT$ are the best-fit blackbody temperatures. The pulsed fractions
(PF) are estimated with the ``max--min PF'' (see the text) in 0.5--2.5\,keV,
0.3--5\,keV, and 0.8--2\,keV for PSRs J1119$-$6127, J1819$-$1458, and
J1718$-$3718, respectively.} \tablerefs{(1) \citealt{rmg+09}; (2)
\citealt{zkm+11}. }
\end{deluxetable*}

\subsubsection{Non-thermal Emission}
The X-ray spectrum of \psr\ clearly shows a PL component with a photon index
of $\Gamma=2.1\pm0.8$ and an unabsorbed flux of $5^{+6}_{-2}\times
10^{-14}$\ergcm\ in the 0.5--8\,keV range. Since the RCS model gives a poor
fit, this seems unlikely to be up-scattering of thermal photons as in
magnetars. We believe that it could be synchrotron radiation originated from
the PWN or from the pulsar magnetosphere. Assuming constant surface
brightness, the PWN would provide an unabsorbed flux of only $8\times
10^{-16}$\ergcm\ in the 0.5--8\,keV range in the central 2\farcs5-radius
region. However, the actual PWN contribution is likely to be higher, since
compact PWN emission is often found to be peaked toward the pulsar
\citep[see][]{nr08}. Deeper \cxo\ observations are needed to determine if
the PL photon indices of the pulsar and PWN components are consistent.
In addition to PWN emission, young rotation-powered pulsars generally show
strong magnetospheric emission in X-rays \citep[e.g.][]{nsg+08}. However, this
emission is expected to be highly pulsed \citep{zc02}, which is not observed
in our case. At a distance of 8.4\,kpc, the non-thermal emission we found has
a luminosity of $L_X=4.2\times 10^{32}$\ergs\ (0.5--8\,keV), a few times larger
than the flux $8.5\times10^{31}$\ergs\ suggested by \citet{zj05} based on the
outergap model.

\subsubsection{Connection with Other High-$B$ Pulsars and Magnetars}
Among all high-$B$ radio pulsars listed in \citet{nk11}, only four (PSRs
J1119$-$6127, J1819$-$1458, J1718$-$3718, and J1846$-$0258) are bright enough
to have X-ray pulsations detected. Except PSR J1846$-$0258, which has a purely
non-thermal spectrum \citep{gvb+00,nsg+08}, the rest show thermal emission
with high BB temperature around 0.1--0.2\,keV. We summarize their timing and
spectral properties in Table~\ref{table:highbpsr}. Interestingly, their X-ray
pulse profiles are all single-peaked and well aligned with the radio pulse,
and show large modulations with PFs from 40\% to 60\%. These striking
similarities suggest that the physics we discussed above in
Section~\ref{sect:psrthermal} could be applicable to other high-$B$ pulsars. More
X-ray observations are needed to confirm this idea. In particular, PSRs
J1734$-$3333 and B1916+14, which are detected in X-rays but for which no
pulsations have yet been found \citep{ozv+12,zkg+09,okl+10}, would be ideal
targets.

PSR J1846$-$0258 is a very special object in the class, as it has shown
magnetar-like activity, including short X-ray bursts \citep{ggg+08}
accompanied by a glitch with unusual recovery properties \citep{lkg10,lnk+11}.
This led to speculation that high-$B$ radio pulsars could be magnetars in
quiescence. It has been proposed that radio pulsars and magnetars could be a
unified class of objects \citep{km05,pp11a} and high-$B$ pulsars could exhibit
magnetar-like bursting behavior, albeit less frequently ($\sim$1 per century).
Radio timing of \psr\ revealed a glitch in 2007 that induced abnormal radio
emission \citep{wje11}. It is unclear if the pulsar exhibited any X-ray
variability at the same time, due to the lack of sensitive X-ray monitoring.
Our \xmm\ observations in 2011 are not useful, since the post-outburst flux
relaxation timescale is likely shorter than a few years
\citep[e.g.][]{lnk+11}. This highlights the importance of X-ray monitoring or
rapid post-glitch follow-up in the study of high-$B$ radio pulsars and
magnetars.

\subsection{SNR \snr}
\subsubsection{SNR Environment and Evolution}
We have carried out spectral analysis of three brightest regions in SNR \snr.
The spectra are well-described by absorbed two-component models. While the
best-fit absorption column densities of the pulsar and region 1 are
consistent ($\sim2\times 10^{22}\,\mathrm{cm}^{-2}$), the values are lower for
regions 2 and 3 ($\sim1\times 10^{22}\,\mathrm{cm}^{-2}$). These seem to
suggest an east-west gradient of \nh\ across the field. This is more evident
in the three-color image in Figure~\ref{fig:img}: the western part of the SNR
is redder than the eastern part, indicating a softer spectrum which could be
due to a lower absorption. This trend is consistent with previous findings
\citep{pkc+01,gs05,gkc+05,sk08}, and can be attributed to a dark molecular
cloud DC 292.3$-$0.4 to the east of the pulsar as first proposed by
\citet{pkc+01}.

The SNR emission consists of a soft thermal component with $kT<1$\,keV and a
hard component that can be described either by a non-thermal PL of photon
index $\Gamma=2.9-3.5$ or by high temperature non-equilibrium ionization
plasma with $kT=1-2$\,keV. While non-thermal synchrotron emission is often
observed in young SNRs, the PL photon index we found ($\Gamma\sim3.5$) is
higher than the typical values of 2--3 in other systems \citep[see a review
by][]{rey08}. In addition, the fit statistics are slightly worse than that of
the two-temperature model. These seem to suggest that the latter may be a more
plausible model. In this case, the solar abundances of the high-temperature
plasma suggest a circumstellar or interstellar origin, hence, the reverse
shock probably has not yet propagated through the ejecta. Together with the
high temperature and small ionization timescales, these provide support for
the young age of the system. The emission measure could give us a handle on
the ambient density $n_0$. Assuming a slab geometry for the emitting regions
with thickness equal to the width, and taking an electron-to-proton density
ratio of 1.2 and a shock compression ratio of 4, we obtained $n_0\sim
0.1$\,cm$^{-3}$ in all three regions. The low-temperature component, however,
is more difficult to interpret. Following the same calculation, we estimate
$n_0$ of 0.6, 0.03, and 0.07\,cm$^{-3}$ in regions 1, 2, and 3, respectively.
This could suggest dense clump for region 1. For regions 2 and 3, a comparison
to the large ionization timescales ($\tau \gtrsim 10^{12}$\,s\,cm$^{-3}$)
indicates that the plasma could have been shocked $\sim10^5$\,yr ago, which
seems difficult to reconcile with the SNR emission. As we suggest below, the
emission may arise from the progenitor wind interaction with the surrounding.

At a distance of 8.4\,kpc, the 8\farcm5-radius SNR shell has a physical size
of 21\,pc \citep{cmc04}. Such a large radius at a young age requires a high
average expansion rate of $v_s=21\,\mathrm{pc /1.9\,kyr}=1.1\times 10^4$\kms.
If this is the present-day shock velocity, then the proton temperature
would be $kT_p=0.11m_{\rm H}v_s^2= 130$\,keV, where $m_{\rm H}$ is the proton
mass, and an adiabatic index of $\gamma=5/3$ and cosmic abundances (i.e., 90\%
H and 10\% He) are assumed. A comparison to the best-fit $kT\approx 1$\,keV
implies an electron-to-proton temperature ratio of $\sim0.01$, lower than in
most SNRs observed \citep[see][]{vhm+08}. If the ambient medium has a uniform
density of $n_0\approx0.1\,\mathrm{cm}^{-3}$ as estimated above, then the
swept-up mass would be $\sim100\,M_\odot$, much larger than the typical ejecta
mass of $\lesssim10M_\odot$. Hence, the remnant could be in a transition to
the adiabatic phase. Based on the Sedov solution, \citet{cgk+01} derived a
constraint on the ratio between $n_0$ (in cm$^{-3}$) and the explosion energy
$E_{51}$ in units of $10^{51}$\,erg. The updated distance and age estimates
from \citet{cmc04} and \citet{wje11} imply $E_{51}/n_0=200$, requiring an
exceptionally large explosion energy or a very low ambient density. It has
been proposed that a newborn neutron star with magnetar-like $B$-field
($\sim10^{15}$\,G) and a short spin period ($\sim1$\,ms) would spin down
quickly via magnetic braking. As a result, the huge amount of rotational
energy released ($\sim10^{52}$\,erg) could accelerate the SNR expansion
\citep{ah04}. While we cannot rule out this possibility, the explosion
energies of three SNRs associated with magnetars (Kes 73, CTB 109, and N49)
are found to be close to the canonical value of $10^{51}$\,erg \citep{vk06},
which is not consistent with this hypothesis.

As an alternative, the supernova could have occurred in a low-density wind
bubble \citep[e.g.,][]{gbm+99,gva06,nrb+07}. This scenario was briefly
mentioned by \citet{cmc04} and here we further explore this idea.
\citet{che05} discussed different types of core collapse supernovae and the
properties of their remnants. Type IIP and IIL/b supernovae are the end
products of red supergiants (RSGs), which have initial masses
$\sim9-25M_\odot$ and $\sim25-35M_\odot$, respectively \citep{hfw+03},
and they are surrounded by dense circumstellar RSG winds extending to a few
pc. In contrast, supernovae of Type Ib/c are believed to originate from
Wolf-Rayet (WR) progenitors, which are massive stars $\gtrsim35M_\odot$ with
high mass-loss rates and high-velocity winds, which are about two orders of
magnitudes faster than RSG winds. For a WR star evolved from an RSG, the fast
WR wind would sweep up the slow RSG wind ejected earlier to evacuate a
low-density wind bubble of radius $\gtrsim10$\,pc over the lifetime
$\sim2\times10^5$\,yr of the WR phase \citep{che05}. The swept-up shell is
characterized by clumpy structure due to Rayleigh-Taylor instabilities, and
overabundance in N and underabundance in C and O \citep{glm96}.

We argue that SNR \snr\ is unlikely to be a Type IIP or IIL/b supernova
because in these cases the remnant should expand into the dense circumstellar
RSG winds. A Type Ib/c event with a WR progenitor, one of the possibilities
mentioned by \citet{che05}, therefore seems more likely. In this picture, the
SNR had a long free-expansion phase in the low-density wind bubble, with the
initial expansion rate up to a few times $10^4$\kms, as seen in SN 1987A
\citep{gms+97,ngs+08}. When the shock reached the cavity boundary at large
radius, it began to encounter the clumpy swept-up shell and hence decelerated.
The shock interaction then gave rise to the observed hard component in the
spectra, either thermal or non-thermal. This scenario may be able to explain
the soft component too. The large ionization timescales of the emission are
comparable to the lifetime of the WR phase, suggesting that it may
originate from the swept-up RSG wind. Two WR bubbles have been detected in
X-rays and their spectra can be fitted by a two-temperature thermal model with
$kT\sim0.1$ and 1\,keV \citep{wcm+05,tgc+12}. The high-temperature component
is comparable to what we observed for SNR \snr, while the low-temperature
component may be too absorbed to detect in our case. In addition, hydrodynamic
simulations and observations suggest a density of $\sim0.1$\,cm$^{-3}$ in the
hot bubbles \citep{glm96,dwa07b,tgc+12}, which is also consistent with our
findings. Deeper X-ray observations in the future can probe the chemical
composition of the circumstellar medium to confirm its nature. Moreover, if
the SNR is detected in optical wavelengths, then spectral line observations
could directly measure the shock velocity to reveal the evolutionary state of
the remnant. 

\subsubsection{Progenitors of High-$B$ Pulsars and Magnetars}
In addition to \psr, only two other young high-$B$ radio pulsars have
associated SNRs detected: PSR J1846$-$0258 in Kes 75 and PSR B1509$-$58 in
MSH 15$-$5\emph{2}. Intriguingly, both of them were claimed to have a WR
progenitor \citep{msb+07,gbm+99}. There is also observational evidence hinting
at a similar mass range for magnetar progenitors, e.g., 1E 1048.1$-$5937, CXOU
J1647$-$4552, and SGR 1806$-$20 \citep{gmo+05,mcc+06,fng+05}, although
intermediate masses were suggested for two cases: 1E 1841$-$045 and SGR
1900+14 \citep{che05,dfk+09}. This raises the important question regarding a
possible correlation between progenitor mass and neutron star magnetic field.
Recently, \citet{che11} compared different classes of young neutron stars in
SNRs, including magnetars, radio pulsars, and central compact objects, and
suggested a tendency for higher $B$-field objects to have more massive
progenitors. However, it was noted that other parameters, such as rotation,
metallicity, and binarity, likely also play an important role.

The origin of magnetic fields in neutron stars has long been an open problem.
Theories suggest that it could be the fossil $B$-fields of the progenitors
preserved during core collapse \citep[e.g.,][]{wol64,rud72,fw06} or from field
amplification via an $\alpha$--$\Omega$ dynamo in rapidly rotating stars
\citep[e.g.,][]{dt92,td93}. A positive correlation between progenitor mass and
neutron star $B$-field may be expected in both scenarios \citep{fw08}. Massive
stars tend to have higher surface $B$-fields \citep{pwd+08}. They also spend
less time in the hydrogen and helium burning phases, during which significant
braking occurs. Hence, the core could retain a large angular momentum,
resulting in efficient dynamo action. If further studies confirm that the
progenitors of high-$B$ radio pulsars and magnetars have comparable mass, then
no matter which of the above mechanisms is at work, the $B$-fields in these
objects could be generated in a similar way. In this respect, magnetars could
have the same formation channel as radio pulsars, which would support a
unification of these classes of objects \citep{km05,kas10,pp11a}.

\section{Conclusion}
We performed a detailed X-ray study of the young high-$B$ radio pulsar
J1119$-$6127 and its associated SNR \snr\ using deep \xmm\ and \cxo\
observations. The pulsar emission exhibits strong pulsations below 2.5\,keV,
with a single-peaked profile that aligns with the radio pulse. Such a
single-peaked profile and alignment seem common among thermally emitting
high-$B$ pulsars, and we showed that the observed pulsed profile can be
modeled by a large hot spot near the magnetic pole. The pulsar spectrum is
well fitted by an absorbed BB plus PL model. The BB temperature
$kT=0.21\pm0.04$\,keV is highest among young radio pulsars. The thermal
emission has a bolometric luminosity of $1.9^{+1.9}_{-0.8}\times
10^{33}$\ergs, which is consistent with neutron star cooling and no heating by
$B$-field decay is needed. However, passive magnetic effects, including
anisotropic heat conduction and beaming, must play a part in the high
temperature and large modulation of the emission.

The spectra of SNR \snr\ are best fitted by a two-component model with solar
abundances. The soft component has a thermal origin with $kT\lesssim1$\,keV
and a large ionization timescale, while the hard component can be described
either by a PL with $\Gamma=2.9-3.5$ or by higher temperature
($kT=1-2$\,keV) thermal emission with low ionization. For a young age of
1900\,yr inferred from the pulsar's spin down, the 21\,pc SNR shell diameter
(at a distance of 8.4\,kpc) implies a fast expansion in the past. This could be
the result of exceptionally large explosion energy $\gtrsim 10^{52}$\,erg or
supernova in a wind cavity.  We believe that the latter is more plausible, and
this could suggest a Type I b/c supernova with a WR progenitor.
WR stars have a typical mass $\gtrsim35\,M_\odot$ and they have also
been proposed as the progenitors for two other high-$B$ radio pulsars. There
is evidence that magnetar progenitors could be in a similar mass range,
leading to a speculation that the two classes of neutron stars may have the
same formation channel. If confirmed, it will provide a strong support to the
idea of unification of these classes of objects. 

We note that late in the preparation of this manuscript, we became aware of
work \citep{ksg12} on the study of the SNR using the same archival \cxo\ data
and a subset ($\sim50$\,ks) of \xmm\ data we have used in our study. While our
deeper observations gave different spectral parameters for SNR \snr, we
arrived at similar conclusions regarding a possible massive progenitor for
\psr.  

{\it Facilities:}
\facility{\emph{CXO} (ACIS)}
\facility{\emph{XMM} (EPIC)}

\acknowledgements
We thank Bryan Gaensler for providing the radio image, and we thank Pat Slane
and Daniel Castro for useful discussions. We also thank the referee for useful
comments. This work was based on observations obtained with \emph{XMM-Newton},
an ESA science mission with instruments and contributions directly funded by
ESA Member States and NASA. This research has made use of the NASA
Astrophysics Data System (ADS) and software provided by the Chandra X-ray
Center (CXC) in the application package CIAO. V.M.K.\ holds the Lorne Trottier
Chair in Astrophysics and Cosmology and a Canadian Research Chair in
Observational Astrophysics. This work is supported by NSERC via a Discovery
Grant, by FQRNT via the Centre de Recherche Astrophysique du Quebec, by CIFAR,
and a Killam Research Fellowship. W.C.G.H.\ appreciates the use of the
computer facilities at the Kavli Institute for Particle Astrophysics and
Cosmology, and acknowledges support from the Science and Technology Facilities
Council (STFC) in the United Kingdom.

\end{document}